\documentclass[aps,prd,reprint,superscriptaddress,nofootinbib]{revtex4-1}

\usepackage{xcolor}
\usepackage{amssymb}
\usepackage[utf8]{inputenc}
\usepackage{amsmath}
\usepackage{graphicx}
\usepackage{bbm}

\usepackage[colorlinks,pdfusetitle]{hyperref}

\newcommand{\orcid}[1]{\href{https://orcid.org/#1}{#1}}

\begin{document}

\title{Inflation meets neutrinos}

\author{Gabriela Barenboim}
\email{Gabriela.Barenboim@uv.es}
\thanks{\orcid{0000-0002-3249-7467}}
\affiliation{Departament de Fisica Te\`orica and IFIC, Universitat de Val\`encia, 46100 Burjassot, Spain}

\author{Peter B.~Denton}
\email{pdenton@bnl.gov}
\thanks{\orcid{0000-0002-5209-872X}}
\affiliation{Physics Department, Brookhaven National Laboratory, Upton, New York 11973, USA}

\author{Isabel M.~Oldengott}
\email{Isabel.Oldengott@uv.es}
\thanks{\orcid{0000-0002-2089-3112}}
\affiliation{Departament de Fisica Te\`orica and IFIC, Universitat de Val\`encia, 46100 Burjassot, Spain}

\date{March 5, 2019}

\begin{abstract}
Constraints on inflationary models typically assume only the standard models of cosmology and particle physics.
By extending the neutrino sector to include a new interaction with a light scalar mediator ($m_{\phi}\sim$MeV), it is possible to relax these constraints, in particular via opening up regions of the parameter space of the spectral index $n_s$.
These new interactions can be probed at IceCube via interactions of astrophysical neutrinos with the Cosmic Neutrino Background for nearly all of the relevant parameter space.
\end{abstract}

\maketitle

\section{Introduction}
\label{Introduction}
The success of the the Big Bang Theory is evident due to the incredible agreement between the model and the anisotropies in the Cosmic Microwave Background (CMB).
In order to explain certain questions such as the homogeneity of the universe on large scales, the flatness of the universe, and the lack of magnetic monopoles, some model of inflation is usually assumed.
Since directly probing the details of inflation is difficult, a plethora of models exist on the market.
Due to recent precision measurements from Planck \cite{Ade:2015xua} many of these models have become ruled out due in large part to constraints of the spectral index $n_s$ and the tensor-to-scalar ratio $r$.
It has since been shown that some regions of $n_s$ parameter space can be allowed by including new neutrino interactions \cite{Oldengott:2017fhy,Lancaster:2017ksf,Kreisch:2019yzn}.

In this paper we explore the neutrino phenomenology involved in opening up the parameter space to models of inflation previously thought to be ruled out.
In particular, this new interaction will cause high energy ($\sim$TeV-PeV) neutrinos traveling astrophysical distances to resonantly scatter and lose energy off the Cosmic Neutrino Background (C$\nu$B).
By looking for dips in the spectrum, experiments like IceCube and KM3NeT can discover or constrain these models.

In section \ref{sec:NSI} we discuss the nature of the new neutrino interaction.
We then look at how cosmology is modified by such a new interaction in section \ref{sec:cosmology} and derive the allowed region in parameter space by performing a Markov Chain Monte Carlo (MCMC) analysis.
In section \ref{sec:Consequences for inflation} we discuss how different inflation models are affected.
Finally, in section \ref{sec:IceCube} we present an overview of how the new interaction could be directly probed, and we conclude in section \ref{sec:conclusions}.

\section{Non-standard neutrino interactions}
\label{sec:NSI}
The Standard Model (SM) of particle physics assumes neutrinos to be exactly massless. The discovery of neutrino oscillations \cite{Fukuda:1998mi,Ahmad:2002jz} has therefore been a clear hint to the existence of physics beyond the SM. Almost every attempt to account for neutrino masses necessitates the existence of yet unobserved particles or yet unobserved interactions in the neutrino sector. 

The simple extension of the SM by right handed Dirac neutrinos however imposes a hierarchy problem, as there is no reason why neutrinos should be more than six orders of magnitude lighter than the charged leptons. If neutrinos are Majorana particles and $U(1)_{B-L}$ is broken, the see-saw mechanism \cite{Minkowski:1977sc} provides an easy way out: Since the mass of the right-handed sterile state is not induced by the Higgs mechanism, the right-handed neutrino could be much heavier than their left-handed active partners. After diagonalizing the mass matrix the active neutrino states would naturally acquire small masses. 

This being said, there is also the possibility that $U(1)_{B-L}$ is spontaneously broken. In that case, we expect the existence of a new Goldstone particle -- the Majoron -- that couples to neutrinos via Yukawa coupling, 
\begin{equation}
\mathcal{L}= g_{\alpha \beta} \bar{\nu}_{\alpha} \nu_{\beta} \phi\,,
\label{Lagrangian}
\end{equation}
where $\alpha $ and $\beta $ stand for flavour or mass eigenstates while $\phi$ is flavour blind

Therefore, we expect the appearance of non-standard neutrino interactions in those models. However, in general this kind of interaction is by no means limited to the existence of a Majoron. The parameterization above remains agnostic about the precise nature of the scalar particle which could be linked to dark sectors or dark matter.

In the following, we restrict our discussion to diagonal couplings, i.e.~$g_{\alpha \beta} \equiv g\mathbbm1_3$ and therefore $g$ has the same form in both flavour and mass basis. The coupling $g$ can be constrained in three kinematic regimes: when the scalar mass $m_{\phi}$ is i) much smaller, ii) comparable or iii) much larger than the center of mass energy $\sqrt{s}$. A comprehensive overview over the constraints as well as their ranges of validity in the $(g,m_{\phi})$-plane can be found in \cite{Ng:2014pca}. Those constraints on $g$ are obtained from different observations: super novae neutrinos \cite{Kolb:1987qy,Brune:2018sab}, big bang nucleosynthesis (BBN) \cite{Ahlgren:2013wba} and the decay of the $Z$ boson \cite{Bilenky:1992xn,Laha:2013xua} help to constrain interactions of the form \eqref{Lagrangian}. A relatively large parameter range of $(g,m_{\phi})$ is however still allowed. Measurements of the cosmic microwave background (CMB) can tighten the bounds, but interestingly also point out a parameter range of $(g,m_{\phi})$ that is in agreement with all observations while neutrinos self-interact according to equation \eqref{Lagrangian}.

\section{A hint from cosmology}
\label{sec:cosmology}
Cosmological observations have proven to be a powerful tool in order to constrain physics beyond the standard model at energies that are out of reach in laboratory experiments. One of the most famous examples thereof is the constraint on the sum of neutrino masses from measurements of the CMB in combinations with Baryonic Acoustic Oscillations (BAO), $\sum m_{\nu} < 0.12$ eV (95\% CL) \cite{Aghanim:2018eyx}. This bound assumes a thermal distribution of cosmic neutrinos and can be 
relaxed if the average momentum of the CMB neutrinos is larger than that of a perfectly thermal distribution \cite{Oldengott:2019lke}.
Additionally, it has been demonstrated in \cite{Oldengott:2017fhy,Kreisch:2019yzn,Lancaster:2017ksf,Cyr-Racine:2013jua,Archidiacono:2013dua} that the CMB can also constrain models of non-standard neutrino interactions, like the Majoron models described in the previous section. 

\subsection{Impact of non-standard neutrino interactions on cosmological observables}
Let us explain the impact of non-standard interactions on the CMB in a bit more detail in the following. According to the SM neutrinos decouple from the cosmic plasma at around $T \approx 1$ MeV. Therefore, at the time of recombination, i.e.~$T \approx 0.3$ eV, neutrinos are usually assumed to be entirely free-streaming. Any form of non-standard interactions in the neutrino sector is expected to change the free-streaming behavior of neutrinos. For the model (\ref{Lagrangian}) and depending on the Majoron mass $m_{\phi}$, this can happen according to two very different thermal histories: If $m_{\phi}>\sqrt{s}$ at all relevant time scales, the population of the Majoron itself would be thermally suppressed such that the Lagrangian (\ref{Lagrangian}) only introduces non-standard neutrino self-interactions. In such a scenario neutrinos would decouple from the cosmic plasma at the standard weak decoupling time, but remain coupled to each other until eventually much later times, when the Hubble rate ($\sim T^2/m_{\mathrm{Pl}}$) overtakes the neutrino self-interaction rate ($\sim g^4 T_{\nu}^5/m_{\phi}^4$). By contrast, if the Majoron is effectively massless at the relevant time scales ($m_{\phi}<\sqrt{s}$), neutrinos would decouple at the standard weak decoupling time, free-stream for some time and \textit{recouple} at later times when the Hubble rate overtakes the neutrino interaction ($\sim g^4 T$) rate. The time of recoupling would also mark the time when the Majoron gets produced. We focus on the first scenario (the delayed decoupling scenario) in the rest of this work. This restricts the validity of our constraints to scalar masses larger than a few hundred keV, in order to ensure that the scalars are non-relativistic at all time scales relevant for the CMB. 

We can understand the cosmological impact of such neutrino self-interactions in the following way: Neutrino free-streaming leads to a suppression of the neutrino energy contrast as it transfers power to the anisotropic stress (and to higher multipole moments in the neutrino Boltzmann hierarchy). Therefore, any non-standard neutrino interactions which are effective after the weak decoupling temperature suppress free-streaming and enhance the neutrino energy contrast. 

The formalism that allows to include neutrino self-interactions in CMB calculations has been derived in \cite{Oldengott:2014qra}. In a subsequent paper \cite{Oldengott:2017fhy}, it has been shown that neutrino self-interactions mediated by a massive Majoron (\ref{Lagrangian}) lead to a scale-dependent enhancement of the anisotropy spectrum of the CMB. 

As shown in \cite{Kreisch:2019yzn}, the impact on the matter power spectrum is mainly due to an increased amplitude of the gravitational potential at horizon entry. This results in a suppression of the matter power spectrum at small wavelengths and to a boost at wavelengths entering the horizon at the time of neutrino decoupling.

\subsection{MCMC analysis}

Since the impact of neutrino self-interactions on the CMB is degenerate in the coupling $g$ and the Majoron mass $m_{\phi}$, it is convenient to introduce an effective four-point coupling (in analogy to the Fermi coupling)
\begin{equation}
G_{\mathrm{eff}}=\frac{g^2}{m_{\phi}^2}\,.
\label{Geff}
\end{equation}

Measurements of the CMB temperature and polarization anisotropy spectra can be used in order to constrain this effective neutrino coupling $G_{\mathrm{eff}}$ (\ref{Geff}). Such analyses have been performed in \cite{Oldengott:2017fhy,Lancaster:2017ksf,Kreisch:2019yzn} with Planck 2015 data \cite{Ade:2015xua} and in \cite{Cyr-Racine:2013jua,Archidiacono:2013dua} with Planck 2013 data \cite{Ade:2013zuv}. Interestingly, the analyses reveal a bimodal posterior distribution in the effective coupling. As expected, the major mode demands neutrinos to behave not too different from the standard assumption, i.e.~to be almost free-streaming\footnote{Note that even this mode actually only demands neutrino to start free-streaming at about $\approx 20$ eV.}. We refer to this mode as the $\Lambda$CDM mode in the following, as the posterior distributions of all cosmological parameters mainly reflect those of the standard $\Lambda$CDM limit with $G_{\mathrm{eff}}=0$. More remarkably, due to a degeneracy of $G_{\mathrm{eff}}$ and some other cosmological parameters (mainly the sound horizon $\theta_s$ at last scattering and the spectral index $n_{\mathrm{s}}$) there exists another allowed region in the cosmological parameter space which allows neutrinos to have very strong interactions in the ballpark of $G_{\mathrm{eff}} \approx 3 \times 10^{-2} \mathrm{MeV}^{-2}$. An interesting consequence of the interacting neutrino mode is a higher value of the Hubble constant $H_0$. This is an appealing feature, as the CMB provides a 2-3 $\sigma$ \cite{Aghanim:2018eyx} lower value of the Hubble constant than local measurements do \cite{Riess:2016jrr}.
In a 1-parameter extension of $\Lambda$CDM by $G_{\mathrm{eff}}$ this tension is only weakened but not resolved. Adding also $N_{\rm eff}$ and $\sum m_{\nu}$ as free parameters to the anlasyis fully alleviates the Hubble parameter tension, and could be related to light sterile neutrino hints \cite{Kreisch:2019yzn}. We however follow a minimalistic approach and only extend the neutrino sector by one additional parameter, i.e.~the neutrino self-coupling $G_{\mathrm{eff}}$. 

\begin{table}
\centering
\caption{Mean values and limits of the most affected cosmological parameters within the self-interacting neutrino mode and the $\Lambda$CDM mode. Quoted limits are at 95\% confidence limits, except the one marked with a $*$ which is at 68\%.}
\label{table_limits}
\begin{tabular*}{\columnwidth}{@{\extracolsep{\fill}}llll@{}}
\hline
& self-interacting mode & $\Lambda$CDM mode \\
\hline
\vspace{0.2cm}
$100\theta_{\mathrm{s}}$            & $1.0463 ^{+0.0018}_{-0.0028}$       & $1.0421^{+0.0009}_{-0.0009}$          \\
\vspace{0.2cm}
$n_{\mathrm{s}}$              & $0.941^{+0.016}_{-0.017}$       & $0.964^{+0.015}_{-0.016}$         \\
\vspace{0.2cm}
$\log_{10}(\mathrm{G}_{\mathrm{eff}} [\mathrm{MeV^{-2}]})$            & $-1.68^{+0.43}_{\mathbf{-0.13^*}}$       & $<-3.04$      \\
\vspace{0.2cm}
$r$              & $<0.11$     & $<0.11$        \\
\vspace{0.2cm}
$H_0 [\mathrm{km/s/Mpc]}$              & $70.06^{+2.21}_{-2.31}$         & $68.35^{+1.94}_{-1.84}$           \\              
\hline
\end{tabular*}
\end{table}

A further remarkable feature of the self-interacting neutrino mode is the fact that is is accompanied by a lower value of the spectral index $n_{\mathrm{s}}$, namely in the region $n_{\mathrm{s}} \approx 0.94$. 
As we will discuss in detail in the next section, this can have important consequences on the selection of inflationary models. Since inflationary model selection is usually performed in the posterior plane of the spectral index $n_s$ and the tensor-to-scalar ratio $r$ (see e.g.~\cite{Akrami:2018odb}), we extended the work of \cite{Oldengott:2017fhy,Lancaster:2017ksf} by adding $r$ as an additional parameter to the analysis. Therefore, we have two extra parameters in addition to the six cosmological base parameters, i.e.
\begin{multline}
\left\{ \omega_{\mathrm{b}}, \omega_{\mathrm{cdm}}, 100\theta_s, \ln(10^{10}A_{s }), n_s, z_{\mathrm{reio}} \right\} \\ 
+ \log_{\mathrm{10}}(G_{\mathrm{eff}}) +r\,.
\label{cosmo_params}
\end{multline}

\begin{figure*}
\includegraphics[width=\textwidth]{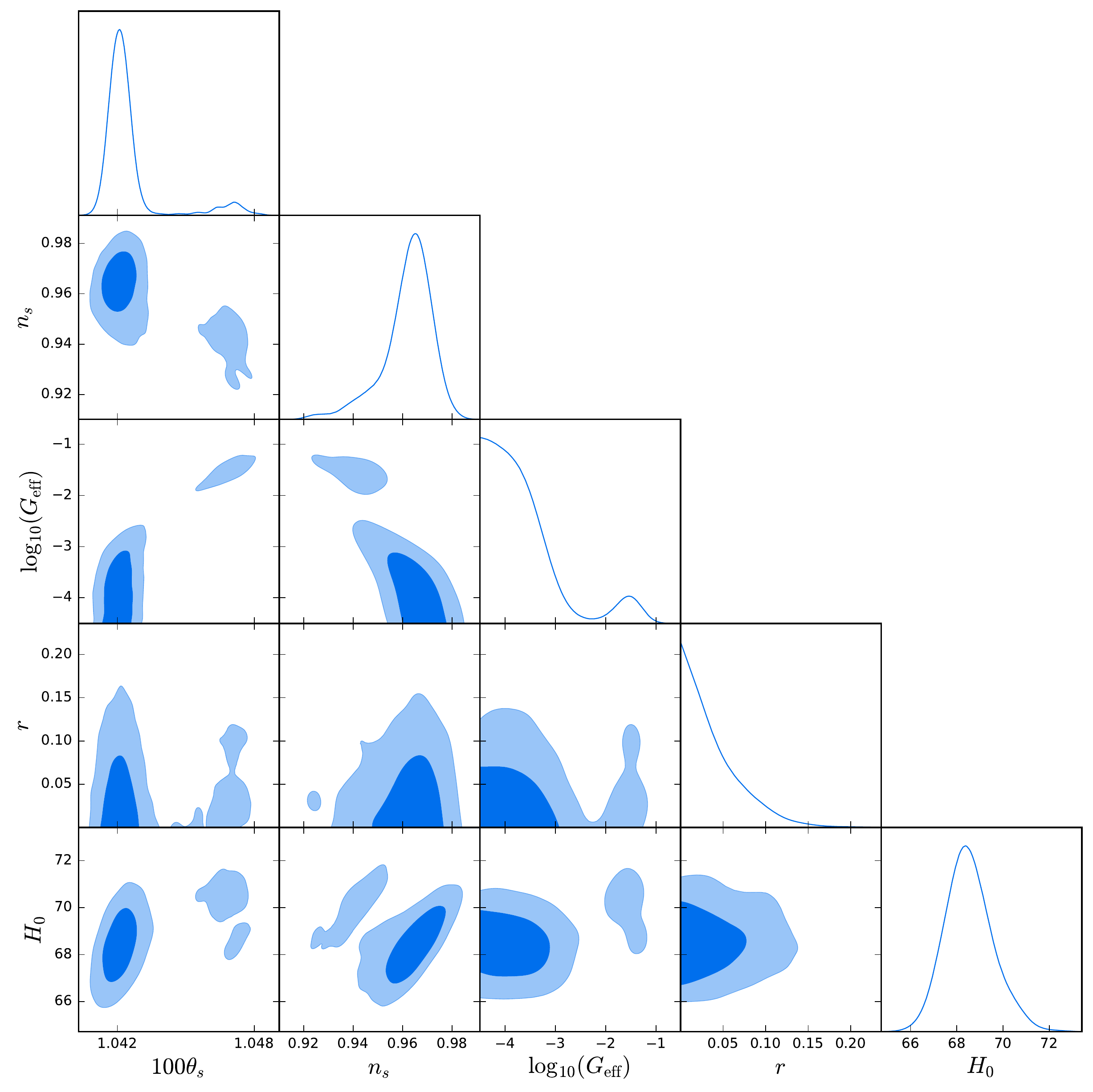}
\caption{2d posterior distributions of the parameters most affected by the effective neutrino coupling $G_{\mathrm{eff}}$.  $G_{\mathrm{eff}}$ is given in units of $\mathrm{MeV}^{-2}$ and $H_0$ in km/s/Mpc.}
\label{triangle} 
\end{figure*}

We modified the numerical Einstein-Boltzmann solver \textsc{class} \cite{Blas:2011rf} in order to take into account the effect of neutrino self-interactions in the neutrino Boltzmann hierarchy. Neutrino interactions also change the propagation of gravitational waves by a suppression of the anisotropic stress \cite{Ghosh:2017jdy}. This also requires a modification of the equations for tensor perturbations in \textsc{class}. Since tensor perturbations only contribute to scales entering the horizon after recombination, this effect however turns out to be negligible. 

Using the Markov Chain Monte Carlo (MCMC) engine Monte Python \cite{Audren:2012wb,Brinckmann:2018cvx}, we explore the cosmological parameter space studying the following combination of data sets:
\begin{description}
\item[\textbf{TT+lowP+lensing}] Temperature anisotropy spectrum and low-$\ell$ polarization plus lensing reconstruction from Planck 2015 \cite{Ade:2015xua}.
\end{description}
This combination of data sets is considered to be a conservative choice according to the Planck collaboration \cite{Ade:2015xua}, since high-$\ell$ polarization data may still be subject to systematic errors.

We adopt the Gelman-Rubin convergence criterion $R<0.01$ and apply flat parameters on all cosmological parameters \eqref{cosmo_params}, restricting the prior range of \linebreak$\log_{10}(G_{\mathrm{eff}}[\mathrm{MeV^{-2}}])$ to $[-4.5,-0.1]$. This choice is justified by the fact that $\log_{10}(G_{\mathrm{eff}}[\mathrm{MeV^{-2}}]) <-4$ results in changes in the CMB temperature and polarization spectra below the percent level. For values \linebreak $\log_{10}(G_{\mathrm{eff}}[\mathrm{MeV^{-2}}])\gtrsim -0.1$ the neutrino Boltzmann hierarchy becomes so stiff that the implicit ODE-solver of \textsc{class} fails to solve it. Neglecting larger values turns out to be a safe assumption as they are still far off the upper limit of the interacting neutrino mode. 

We present our results for the 2D posteriors of the cosmological parameters most affected by the interacting neutrino mode in figure \ref{triangle}. As expected we recover the bimodal posterior distribution for $G_{\mathrm{eff}}$ which has been reported in \cite{Oldengott:2017fhy,Kreisch:2019yzn,Lancaster:2017ksf,Cyr-Racine:2013jua}. We turn our discussion to the posterior in the $(n_{\mathrm{s}},r)$ plane to section \ref{sec:Consequences for inflation}.  

In order to obtain the confidence limits of the two individual modes, we ran two more MCMC analyses: one for the self-interacting mode (which we define by the prior range $\log_{10}(G_{\mathrm{eff}}[\mathrm{MeV^{-2}}])=[-2.5,-0.1]$) and one for the $\Lambda$CDM mode (defined by the prior range $\log_{10}(G_{\mathrm{eff}}[\mathrm{MeV^{-2}}])=[-4.5,-2.5]$). The results of these separate runs can be found in table \ref{table_limits}. Since the two modes are still slightly connected at 95\% confidence limit, we quote the 68\% confidence lower limit in case of the self-interacting mode.

As in previous works \cite{Oldengott:2017fhy,Kreisch:2019yzn,Lancaster:2017ksf}, we find that the $\Lambda$CDM mode is statistically favored over the self-interacting mode, with a difference in the best fit $\chi^2$ values of $\Delta \chi^2=3.4$. Adding polarization or external data such as BAO or direct $H_0$ measurements of course has an impact on the significance of the self-interacting neutrino mode. Based on the extended analysis of \cite{Oldengott:2017fhy,Kreisch:2019yzn,Lancaster:2017ksf} we however do not expect a qualitative change of the bimodal posterior distribution of $\log_{10}(G_{\mathrm{eff}})$ when using different combinations of data sets.

\begin{figure}
\centering
\includegraphics[width=\columnwidth]{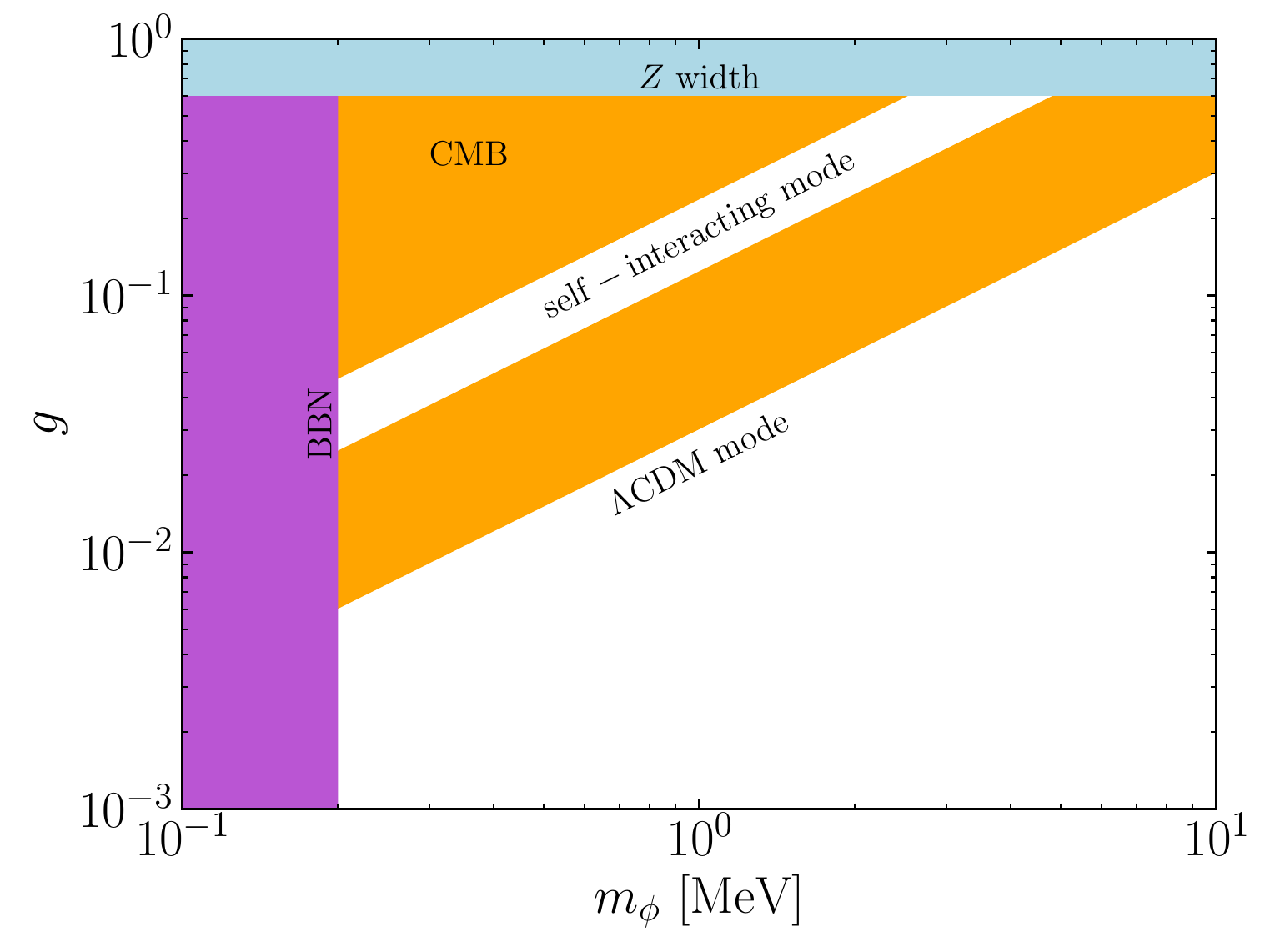}
\caption{Allowed parameter range for $G_{\mathrm{eff}}$ in the $(g,m_{\phi})$-plane. The white region is allowed while the various shaded regions are excluded.
The two disjoint white regions correspond to the two solutions allowed by CMB data: the smaller compact one is that with a new neutrino interaction and the larger one corresponds to the $\Lambda$CDM mode.}
\label{Geff_plot} 
\end{figure}

Figure \ref{Geff_plot} shows how the self-interacting neutrino mode translates into the plane of the coupling $g$ and the scalar mass $m_{\phi}$. We limit the plot to that region of the parameter space  which is still allowed by other observations: Masses below $\sim 0.2$ MeV are excluded by BBN \cite{Ahlgren:2013wba} while couplings larger than $g\sim 0.6$ are excluded by measurements of the decay of the $Z$ boson \cite{Laha:2013xua,Ng:2014pca,Ioka:2014kca}. Realistic values for the scalar mass of the interacting neutrino mode therefore fall into the relatively narrow range of $m_{\phi} \sim [0.2 \, \mathrm{MeV},5 \, \mathrm{MeV}]$. As we discuss in section \ref{sec:IceCube}, this mass range falls by coincidence exactly into the energy window which is testable by IceCube. Let us however first focus on the impact of the self-interacting neutrino mode on constraints of inflationary models. 

\section{Consequences for inflation}
\label{sec:Consequences for inflation}
Despite its indisputable success the Hot Big Bang Model and general relativity do have some 
inconsistencies that can be solved only by a period of accelerated expansion known as inflation.
Inflation is the key to explain the homogeneity, the isotropy, and the flatness of the Universe,
as well as the absence of monopoles. 

However, inflation is not a model but a framework. There are a multitude of inflationary models
in the literature.
And although every inflationary model produces an approximately homogeneous Universe, each does so in somewhat unique ways. Each model of inflation predicts its own small inhomogeneities that essentially behave as a particular models smoking gun. Therefore the experimental observation of inhomogeneities
via CMB anisotropies and structure formation provides a test of the different inflation models.

Inflation models mainly predict two types of perturbations, scalar and tensor, which
turn into density (matter) and gravitational wave fluctuations. Each of them is generally described by a fluctuation amplitude and a dependence on the scale of such an amplitude. In the case of scalar
perturbations the amplitude is called ${\cal P_R}^{1/2}$ and the spectral index $n_s$ while for
gravity waves they are ${\cal P_T}^{1/2}$ and $n_T$, respectively, with the former generally given in terms of
the ratio $r= {\cal P_T}/{\cal P_R}$.

But these four quantities are not independent, only two of them are and as a consequence, theoretical
predictions for the different inflation models as well as data are presented in the $n_s-r$ plane.
This plane is the way to allow or rule out an inflation model. Therefore establishing which region
of this plane is allowed by data is essentially establishing which models of inflation survive the
experimental scrutiny.
 
As we have already seen the inclusion of a new neutrino interaction, completely consistent with
all experimental evidence so far, significantly enlarges the allowed region in the $n_s-r$ plane
and therefore gives a new life to models that would be excluded or under great tension in the absence of such an 
interaction. In the following we will not only show two examples of such models but also discuss how
the interactions capable of giving this second chance to inflationary models can be tested in
neutrino experiments in the near future.

\subsection{Inflation observables}
Specializing the Lagrangian of the form
\begin{equation}
{\cal L}= \frac{1}{2} g^{\mu \nu}\partial_\mu \phi \partial_\nu \phi - V(\phi)\,,
\end{equation}
to the case of a Friedmann-Robertson-Walker metric, 
\begin{equation}
g_{\mu \nu}= \mbox{diag}\{ 1,-a^2(t), -a^2(t),-a^2(t)\}\,,
\end{equation}
results in the equation of motion of the form
\begin{equation}
\ddot{\phi} + 3 H \dot{\phi} + V'(\phi) =0\,,
\end{equation}
where $H=(\dot{a}/a)$ is the Hubble parameter and the prime denotes derivative with respect to $\phi $. The
amplitudes of scalar and tensor perturbations then, are given by
\begin{equation}
A_S^2 = \left.\frac{512 \pi}{75 m_{\mathrm{Pl}}^6} \frac{V^3}{V'^2}\right|_{k=a H} \;\;\;\; ; \;\;\;\; 
A_T^2 = \left.\frac{4}{25 \pi} \frac{H^2}{m_{\mathrm{Pl}}^2}\right|_{k=a H}\,,
\end{equation}
where the above expressions are evaluated at Hubble radius crossing, $k= a H$ with $k$ the comoving
wavenumber and the tensor to scalar ratio is given by
\begin{equation}
r\equiv 16 \; \frac{A_T^2}{A_S^2}\,.
\end{equation}
Defining the spectral indices of scalar and tensor perturbations by
\begin{equation}
n_s -1 \equiv \left.\frac{d \, \mbox{ln} A_S^2}{d \, \mbox{ln} k} \right|_{k= a H} \;\;\;\; ; \;\;\;\;
n_T \equiv \left.\frac{d \, \mbox{ln} A_T^2}{d \, \mbox{ln} k} \right|_{k= a H} \,,
\end{equation}
and imposing the slow-roll regime, i.e.~$\ddot{\phi} \ll 3 H \dot{\phi}$ results in
\begin{equation}
n_s -1 = -6 \epsilon + 2 \eta \;\;\;\; ; \;\;\;\; n_T= -2 \epsilon \,,
\end{equation}
where the slow roll parameters are defined as
\begin{equation}
\epsilon \equiv \frac{m_{\mathrm{Pl}}^2}{16 \pi} \left( \frac{V'}{V}\right)^2 \;\;\;\; ; \;\;\;\;
\eta \equiv \frac{m_{\mathrm{Pl}}^2}{8 \pi} \frac{V''}{V}\,.
\end{equation}
Inflation ends when the field $\phi $ reaches a value such that $\epsilon(\phi_e) = 1 $ and the amount of
inflation generated is quantified by the number of e-folds,
\begin{equation}
N\simeq - \frac{8 \pi}{m_{\mathrm{Pl}}^2} \int_{\phi}^{\phi_e} \frac{V}{V'}\,. 
\end{equation}
The perturbations we observe today were generated 45-60 e-folds before the end of inflation and at this value of the field we have to evaluate the spectral and scalar indices.

It is clear then, that given an inflaton potential, it is straightforward to analyze its phenomenological
imprints on the CMB. Thus, considering the absolute freedom in the potential selection, simple and well motivated potentials are especially welcomed. Among these, two potentials are especially appealing from the particle physics point of view: Natural Inflation (NI) and (small field) Coleman Weinberg (CW) Inflation. Amazingly enough both happen to be either in tension (NI) or ruled
out (CW) if we do not include the possibility of neutrino interactions.

\begin{figure}
\includegraphics[width=\columnwidth]{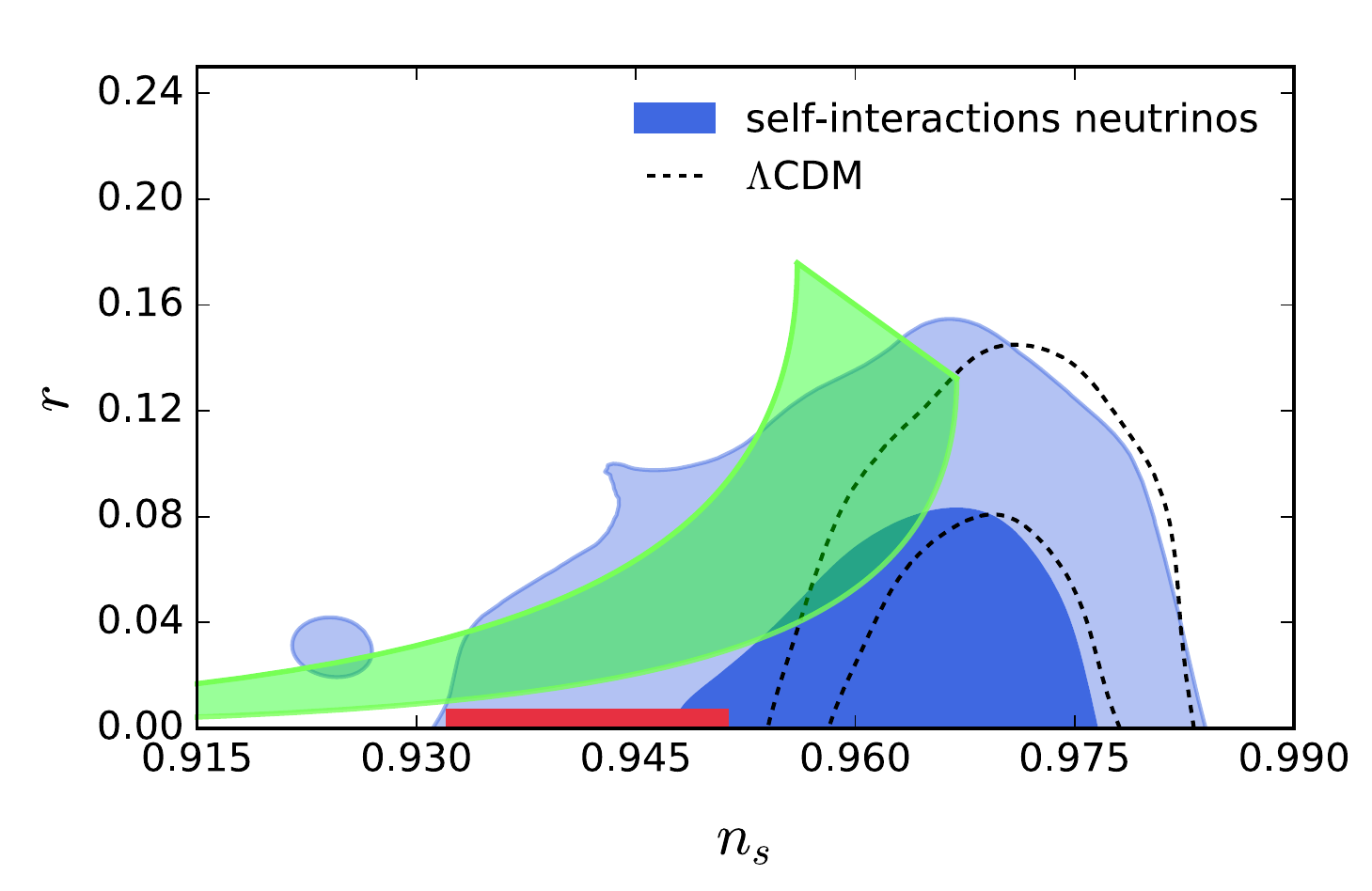}
\caption{Comparison of the predictions of Natural Inflation (green) and Coleman Weinberg inflation (red) with the 68\% (dark blue) and 95\% (light blue) confidence regions for the interacting neutrino mode (same data set as figure \ref{triangle}). Dashed lines mark the confidence regions obtained within the standard $\Lambda$CDM framework assuming free-streaming neutrinos ($G_{\mathrm{eff}}=0$).}
\label{rvsns} 
\end{figure}

\subsection{Inflationary model selection}

Natural Inflation is a technically natural answer to the required flatness of the inflaton potential. 
In the original proposal the inflaton was the Pseudo Nambu Goldstone boson of a broken symmetry, shift symmetry \cite{Freese:1990rb,Adams:1992bn}. The symmetry that precisely explained why the potential was ``nearly" flat. 
Since then many types of candidates have been explored, including hybrid models and multi field models.
From the purposes of this study the precise nature of the inflaton and how it is embedded in a complete
particle physics model is not crucial and therefore we will focus
on the original version of the model, in which there is a unique field rolling down a potential of
the form
\begin{equation}
V(\phi) = \Lambda^4 \left( 1 + \cos(\phi /f) \right)\,,
\end{equation}
and the slow-roll parameters take the form
\begin{align}
\epsilon(\phi) & \simeq \frac{m_{\mathrm{Pl}}^2}{16 \pi \, f^2} \left[ \frac{\sin(\phi/f)}{1+\cos(\phi/f)}\right], \\
\eta(\phi)& \simeq - \frac{m_{\mathrm{Pl}}^2}{16 \pi \, f^2} \,,
\end{align}
while inflation ends when the field take a value such that
\begin{equation}
\cos(\phi_e /f) = \frac{1- 16 \pi (f/m_{\mathrm{Pl}})^2}{1+ 16 \pi (f/m_{\mathrm{Pl}})^2}\,.
\end{equation}
It can be easily seen that for $f< m_{\mathrm{Pl}} $, $n_s $ is essentially independent of the number of e-folds while for $f > m_{\mathrm{Pl}} $, $n_s$ has no dependence of $f$.
The predictions of Natural Inflation can be seen in Figure~\ref{rvsns} for $ 45< N<60 $ (green). We show the posterior distribution in the $(\mathrm{n_s,r})$ plane when neutrinos are allowed to have self-interactions, i.e.~a zoom-in of the $(n_s,r)$ plot in figure \ref{triangle}. As a comparison we also show the posterior contours of the standard case when $G_{\mathrm{eff}}=0$ (dashed) which are consistent with the Planck 2015 results. Although Natural Inflation is not currently ruled out for the $G_{\mathrm{eff}}=0$ case, it is out of the Planck 2015 one-sigma favored region \cite{Freese:2014nla} and expected to be completely excluded if tensor modes are not found at the few \% level. Neutrino interactions keep the model afloat.

Unlike Natural Inflation, Coleman Weinberg potentials do not arise naturally but are unavoidable once 
loop corrections are included in the theory and therefore have been studied extensively \cite{Linde:1981mu,Albrecht:1982mp} (and ruled out
some time ago \cite{Barenboim:2013wra}). A general CW potential evaluated at a renormalization scale $f$ takes the form
\begin{equation}
V(\phi) = A \phi^4 \left[ \ln\left( \frac{\phi}{f}\right) - \frac{1}{4}\right] + \frac{A \, f^4}{4}\,,
\end{equation} 
which gives
\begin{align}
\epsilon(\phi) &= \frac{16 \; m_{\mathrm{Pl}}^2}{\pi \; f^2} \left( \frac{\phi}{f} \right)^6 \ln^2 \left(
\frac{\phi}{f}\right), \nonumber \\
\eta(\phi) &= \frac{ m_{\mathrm{Pl}}^2}{2 \pi \; f^2} \left( \frac{\phi}{f} \right)^2 \left[ \ln \left(
\frac{\phi}{f} \right) + 1\right], \\
N(\phi) &= \frac{2 \pi \; f^2}{m_{\mathrm{Pl}}^2} \left( \mbox{Ei}\left[-2\ln \left(
\frac{\phi}{f} \right)\right] - \mbox{Ei}\left[-2\ln\left(
\frac{\phi_e}{f} \right) \right] \right)\,. \nonumber
\end{align}
From these equations it can be seen that in the small field inflation regime, i.e.~$\phi/f \ll 1 $, $|\epsilon| \ll \eta $ and therefore in CW inflation
\begin{equation}
N \simeq \frac{3}{1-n_s}\,,
\end{equation}
while $r \propto (f/ m_{\mathrm{Pl}})^4 \sim 0 $. The predictions of (small field) CW inflation are shown in Figure~\ref{rvsns} for $ 45< N<60 $. Clearly this model can only survive (and thrive) once neutrino
interactions are included.

The cases presented above are just two examples but convey an important message.
Neutrinos are so far the only evidence of physics beyond the Standard Model we have observed.
We do know already that the possibility of a Majorana mass terms for neutrinos may introduce new scales and interactions not shared by the other fermions.
Right handed neutrinos are singlets under the Standard Model group and therefore could provide a connection to dark matter and the dark sector in a way other fermions cannot.
By ignoring the possibility of neutrino interactions we are missing important regions of cosmological parameter space.
We speculate that many potentially appealing and interesting inflation models were not even considered because they appeared to predict spectral indices that were ruled out in the absence of new neutrino interactions.

Even more, in the following we will show that the existence of these neutrino interactions can be
experimentally tested and searched for in astrophysical experiments that are running now. Thus maybe
in the near future astrophysical experiments can shed a new light on the way we analyze inflation.

\section{Astrophysical neutrino flux}
\label{sec:IceCube}

IceCube has detected the high energy astrophysical neutrino flux for the first time \cite{Aartsen:2013bka} which is now understood to be extragalactic \cite{Denton:2017csz,Aartsen:2017ujz}.
This provides us with a new window to the universe and an opportunity to probe new physics models coupling to neutrinos.
Consistent with our understanding of extreme non-thermal astrophysical phenomenon, the data is largely consistent with a single power law $dN/dE\propto E^{-\gamma}$ with a spectral index $-2.9\lesssim\gamma\lesssim-2.3$.
The astrophysical flux has been measured up to $\mathcal O($few$)$ PeV and down to $\mathcal O($few$)$ TeV.
With the proposed upgrade to IceCube-Gen2 \cite{Ackermann:2017pja} it is likely that the Glashow resonance \cite{Glashow:1960zz} at $E_\nu=6.3$ PeV would be observed and the flux could be measured up to $\sim10$ PeV depending on whether or not there is a cutoff in the flux and what the true spectrum is.
The Glashow resonance occurs at $E_\nu=6.3$ PeV when a high energy $\bar\nu_e$ hits an at rest $e^-$ and creates a $W^-$ on-shell leading to a considerably increased cross section.
Detecting point sources will be easier with IceCube-Gen2 not only because of the increased statistics, but also because the angular resolution of tracks will improve.
Once point sources are found, it may be possible to extend the measurement of the astrophysical flux down to lower energies, possibly $\sim1$ TeV or lower by using the known locations of the sources to cut through the atmospheric background.

\begin{figure}
\centering
\includegraphics[width=\columnwidth]{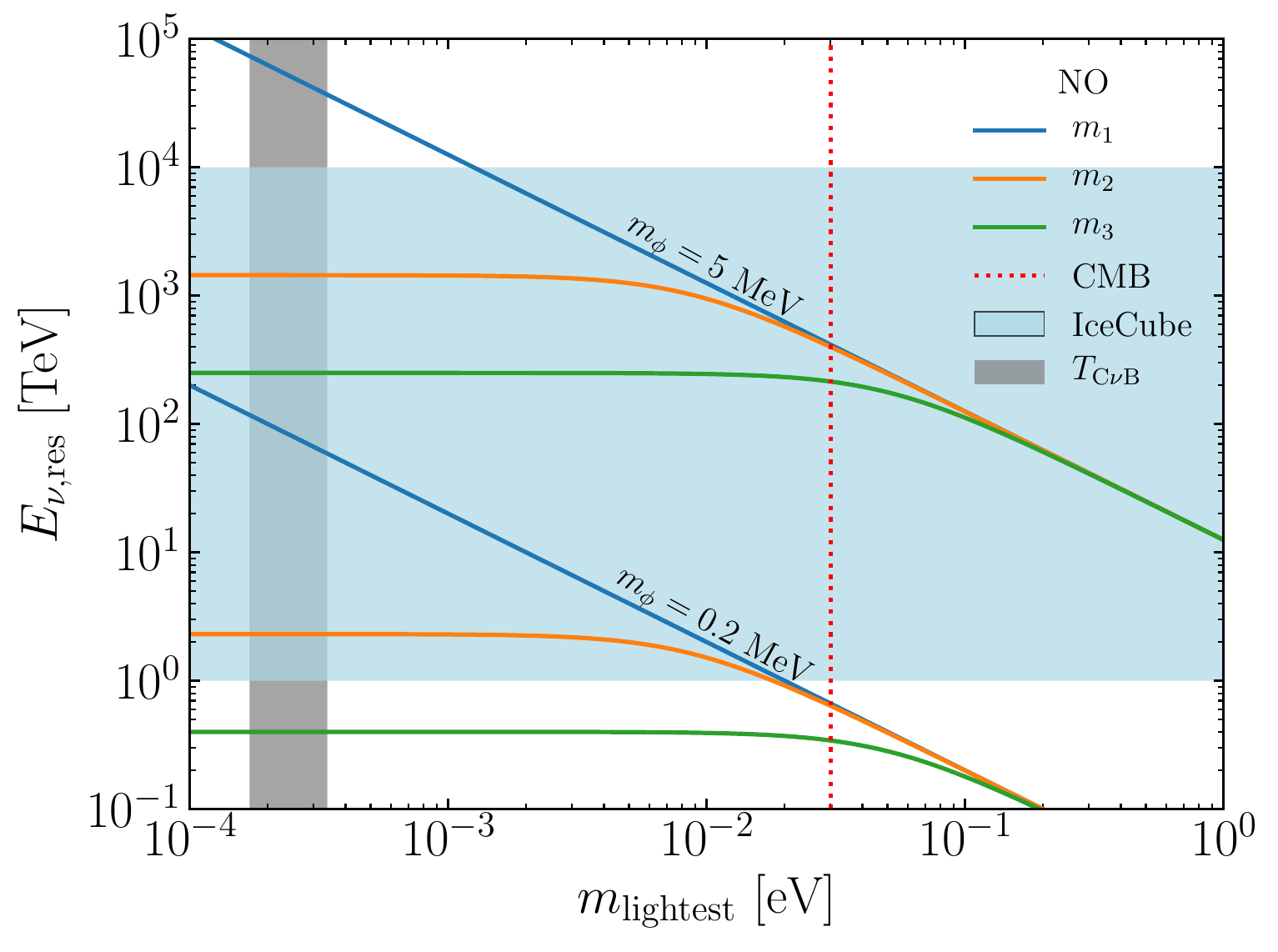}
\caption{The resonant energy of each of the three C$\nu$B mass eigenstates as a function of the lightest neutrino mass in the normal mass ordering.
The upper three curves are for $m_\phi=5$ MeV and and the lower three for $m_\phi=0.2$ MeV.
These curves are for the NO, the IO is generally the same without the middle, orange, $m_2$ curve.
The blue band is the broadest range of energies that IceCube could possibly measure the astrophysical neutrino flux.
The gray band indicates the temperature of the C$\nu$B over $z\in[0,1]$; for $m_{\rm lightest}$ below $T_{\rm C\nu\rm B}$, the resonant energy for $m_1$ levels off as it is relativistic.
The red dashed line indicates the limit from cosmology at $\sum m_\nu<0.12$ eV \cite{Aghanim:2018eyx}.}
\label{fig:resonant IC}
\end{figure}

The presence of such a new mediator $\phi$ given in \eqref{Lagrangian} would lead to a resonant absorption of high energy neutrinos off the cosmic neutrino background (C$\nu$B) \cite{Ng:2014pca,Ioka:2014kca,DiFranzo:2015qea}. 
The resonant energy is
\begin{equation}
E^{\mathrm{res}}_{\nu_i}=\frac{m_\phi^2-m_{\nu_i}^2}{2m_{\nu_i}} \approx \frac{m_{\phi}^2}{2 m_{\nu_i}}\,,
\label{eq:resonant energy}
\end{equation}
for non-relativistic C$\nu$B neutrinos.
Such an absorption would lead to a dramatic signal in IceCube that would be very hard to reproduce with standard astrophysics.
As identified in \cite{DiFranzo:2015qea}, the redshift dependence of the unknown source population will smear out the signal somewhat, but the result is generally independent of the redshift evolution.
While the resonant energy is a function of both $m_\phi$ and $m_\nu$, neither of which are known; they both have some constraints as outlined in fig.~\ref{fig:resonant IC} for the normal mass ordering (currently preferred at $\sim3\sigma$ \cite{deSalas:2017kay,Esteban:2018azc}).
At the upper end of the $m_\phi$ allowable mass range $\sim5$ MeV IceCube will have very good sensitivity as there will be resonances at $E_\nu\gtrsim250$ TeV independent of the absolute mass scale and possibly additional features as well.

As the mass of $m_\phi$ decreases the resonance energy decreases.
At the lower end $m_\phi\sim0.2$ MeV it is the lightest neutrino that will provide the resonance in IceCube's region of interest.
Note that when the lightest neutrino drops below the C$\nu$B temperature the resonance energy becomes independent of the lightest neutrino mass as it becomes relativistic meaning that the blue curves should level out to the left of the gray band.
In this case, unless the lightest neutrino is near the limit from cosmology, IceCube will have sensitivity to this case as well.

While identifying such a dip in the IceCube spectrum does not guarantee that the neutrinos are scattering off the C$\nu$B with such a model, it is an indication of some kind of new physics.
Given a broad enough energy scan it could be possible to identify multiple peaks which would provide an indication that these dips were the result of scattering off the C$\nu$B.
There is currently a hint of a dip at $\sim500$ TeV that has persisted for several years \cite{taboada_ignacio_2018_1286919}, although its significance is quite low.
Up to small factors this roughly corresponds to $m_\phi\simeq5$ MeV for $m_{\rm lightest}$ near the upper limit, and $m_\phi\simeq0.5$ MeV in the limit where $m_{\rm lightest}$ is relativistic.

It is important to note that the current IceCube analysis must assume some spectrum at the Earth since neutrino energies cannot be measured directly without a prior hypothesis.
Typically that prior takes the form of a single power law\footnote{Another IceCube analysis considered a two-component power-law and found it was not preferred over a single power law \cite{Aartsen:2017mau}.}.
We hope that in the future IceCube performs fits with different functions including those with dips for identifying resonant models.
In addition, possible future measurements of neutrinos at even higher energies up to $E_\nu\sim1$ EeV with Auger \cite{Aab:2015kma}, ANITA \cite{Allison:2018cxu}, ARA \cite{Allison:2015eky}, ARIANNA \cite{Nelles:2015tch}, GRAND \cite{Alvarez-Muniz:2018bhp} and POEMMA \cite{Olinto:2017xbi} could extend the reach to larger values of $m_\phi$ provided sufficient energy resolution.
Also the presence of regeneration, a process wherein high energy neutrinos lose energy as they upscatter the C$\nu$B, could also lead to a distinct signal which could be relevant for cosmogenic neutrinos from the interactions of ultra-high energy cosmic rays off the C$\nu$B scattered down to IceCube energies.

\section{Conclusions}
\label{sec:conclusions}
The precision of cosmological data has moved us from testing the generic predictions of the inflationary paradigm to selecting individual models of inflation.
Inflation is governed exclusively by two parameters, therefore extreme care is demanded when we extract these two numbers from data. 
While in the conventional picture inflation models are becoming somewhat constrained, the presence of new neutrino interactions enlarges the allowed values of these two numbers and therefore enlarges the number of models and changes the selection criteria.

This model also has a clear signature at IceCube and we hope that future analyses of the high energy astrophysical flux will look for dips in the spectrum.
Such a dip would provide strong evidence of the C$\nu$B as well as a moderately constrained measurement on the new mediator at the MeV scale; with additional information on the absolute mass scale and the astrophysical sources it is conceivable that the mediator mass could become well-measured.
Finally, while the current dip in the IceCube data at $\sim500$ TeV is not yet significant, it is a tantalizing hint.

\begin{acknowledgments}
GB and IMO acknowledge financial support by MEC and FEDER (EC) Grants No.~SEV-2014-0398, FIS2015-2245-EXP, FPA2014-54459 and the Generalitat Valenciana under Grant No.~PROME-TEOII/2013/017 and by EU Networks FP10 ITN ELUSIVES (H2020-MSCA-ITN-2015-674896) and INVISIBLES-PLUS (H2020-MSCA-RISE-2015-690575).
PBD acknowledges the United States Department of Energy under Grant Contract desc0012704.
Numerical calculations have been performed using the computing resources from the Tirant cluster of University of Valencia.
\end{acknowledgments}

\bibliography{Literature.bib}

\end{document}